\documentstyle[aps,preprint,tighten]{revtex}

\input{psfig}

\begin{document}

\draft

\title{ \bf
Mn $K$-edge XANES studies of La$_{1-x}$A$_x$MnO$_3$ systems (A = 
Ca, Ba, Pb)
}

\author{F. Bridges,$^1$ C. H. Booth,$^{2}$ M. Anderson,$^1$  G. H. Kwei,$^2$  
J. J. Neumeier$^{3}$, J. Snyder,$^4$ J. Mitchell,$^5$ J. S. Gardner$^{2}$
and E. Brosha$^{2}$} 
\address{$^1$Physics Department, University of California, Santa Cruz, CA 95064}
\address{$^2$Los Alamos National Laboratory, Los Alamos, NM 87545}
\address{$^3$Physics Department, Florida Atlantic University, Boca Raton, FL 33431}
\address{$^4$JPL, California Institute of Technology, Pasadena CA 91109-8099}
\address{$^5$Argonne National Laboratory, Argonne, IL  60439}


\maketitle

\begin{abstract}
We present Mn $K$-edge XANES data for a number of manganite systems as a function
of temperature. The main edge (1s-4p transition) for the Ca substituted samples 
is very sharp, almost
featureless, and shifts uniformly upwards with increasing Ca content. 
The interpretation of this result is controversial because
the lack of structure appears difficult to reconcile with a 
mixture of Mn$^{+3}$ and Mn$^{+4}$ ions. We propose a possible solution
in terms of the extended Mn 4p states and hybridization between 
the Mn3d and O2p states. A small T-dependent feature is present in
the main edge; analysis for the charge ordered (CO) sample suggests
a distortion associated with the CO state that increases below T$_{CO}$. The 
manganite pre-edge structure is quite similar to that for a large number of 
other Mn compounds, with two or three small peaks that are ascribed to 1s-3d
weakly allowed dipole transitions plus possibly a small quadrupole component.
The weak dipole transitions are explained as arising from 
a hybridization of the extended Mn 4p state of the excited atom with an 
odd symmetry combination of Mn 3d states on adjacent Mn atoms.  The first 
pre-edge peak, A$_1$, has a small shift to higher energy with increasing 
valence while the next peak, A$_2$, is nearly independent of dopant 
concentration at 300K. However, for the colossal magnetoresistance (CMR) 
samples the A$_2$ pre-edge peak shifts to a lower energy below the 
ferromagnetic (FM) transition temperature, T$_c$, resulting in 
a decrease in the A$_2$-A$_1$ splitting by $\sim$ 0.4 eV. This indicates a
change in the higher energy 3d bands, most likely the minority spin e$_g$ plus
some change in covalency.  In addition, the amplitudes are temperature 
dependent for the CMR materials, with the change in A$_1$, A$_2$ correlated 
with the change in sample magnetization.
We discuss these results in terms of some of 
the theoretical models that have been proposed and other recent XANES studies.

\end{abstract}


\narrowtext

\section{Introduction}

The La$_{1-x}$A$_x$MnO$_{3+\delta}$ systems exhibit a wide range of different
phenomena depending on the concentration, x, of the divalent substitutional
atom, A (A = Ca, Ba, Sr, Pb, etc.) and the O concentration.
These include ferromagnetism,
antiferromagnetism, charge ordering, a metal-insulator transition, and
large magnetoresistance
effects.\cite{Volger54,Wollan55,Schiffer95,Jonker50,Ramirez96,Chen97}
For x roughly in the range 0.2-0.5, these systems have a ferromagnetic
transition at a transition temperature, T$_c$, a metal/insulator
transition (MI) at T=T$_{MI}$, and a "colossal'' magnetoresistance (CMR) 
which reaches it's maximum at
T$_{MR}$, with T$_{MI}$ $\sim$ T$_{MR}$ $\sim$ T$_c$ in many cases. 
The substitution
of a divalent ion for La$^{+3}$ formally changes the average Mn valence to 3+x
(the Mn valence is +3 in LaMnO$_3$  +4 in CaMnO$_3$), and
is usually thought to introduce holes
into the narrow e$_g$ band of Mn 3d-electrons, which are also hybridized with
O 2p states. It is the changing occupation of this hybridized band with x that
leads to many of the observed properties. Excess O in LaMnO$_{3+\delta}$ 
(actually Mn and La vacancies) or La vacancies
can also increase the formal Mn valence, thereby adding carriers to
the system. 

The coupling between charge and
magnetism has been modeled using the double exchange (DE)
mechanism\cite{Zener51,Anderson55,deGennes60} plus strong electron-phonon 
coupling.\cite{Millis95,Millis96,Roder96} 
In the ferromagnetic metallic (FM) state, well below T$_c$,
the charge carriers are assumed to be highly delocalized (large polarons)
and spread out over several unit cells for CMR samples.

In the paramagnetic (PM) state above T$_c$ for CMR samples, there is a significant
increase in the distortions about the Mn atoms compared to the low temperature
data.\cite{Booth96,Booth98a,Booth98b,Subias97,Billinge96} It is
generally assumed that this distortion is a result of the charge carriers
becoming localized on the Mn atoms. However, mobile holes could also be 
located more on the O atoms as is the case for cuprates.\cite{Nucker88}
Consequently, there might be very little change in the charge localized on the 
Mn atoms above and below T$_c$. This raises the question as to how the energy 
shift of the absorption edge relates to valence and the local environment in 
these materials. Is there a mixture of ``ionic-like" +3 and
+4 states, an average valence, as in a metal where all Mn atoms are equivalent, or
something in between? 

Experimentally\cite{Booth98b,Subias97}, the Mn $K$-edge absorption for the 
Ca substituted manganites is
sharp with relatively little structure, and shifts almost uniformly with dopant
concentration, consistent with an average valence state
of v = 3+x. The sharpness of the edge is suggestive of a transition into
a state that is uniform throughout the sample, and initially we interpreted
this result to mean that all Mn sites have comparable local charge densities.
This is difficult to reconcile with the usual assumption of a mixture of
purely {\it local} ionic Mn$^{+3}$ and Mn$^{+4}$ sites.
For example, below we show explicitly that the observed $K$-edge cannot 
be modeled as a
weighted sum of the edges of the end compounds LaMnO$_3$ and CaMnO$_3$ for
the charge ordered (CO) material with 65\% Ca. Note however that
even with similar charge densities, the d-electron wavefunctions and 
the local environment need not be identical at each 
site\cite{Anisimov97,Mizokawa97}. In addition, Tyson {\it et al.}\cite{Tyson00} 
have investigated the K$_{\beta}$ emission
which probes Mn 3d states through the 3p-1s decay, and report that
these spectra for the substituted manganite materials can be modeled as a
weighted sum of the end compounds although the shifts with valence are small.


The pre-edge structure for the Mn $K$-edge consists of 2-3 small peaks 
labeled A$_1$-A$_3$ which have Mn 3d character. These features are observed for
all the transition metals and are generally ascribed to  mixture of
1s-3d quadrupole and  1s-p dipole transitions (made weakly allowed by a hybridization
between 3d states and p-states). Although the latter are assumed to be
dominant, the interpretation of the A$_i$ peaks is still controversial\cite{Joly99}. 
Two important issues are: 1) How large is the quadrupole
contribution and when is it important?  and 2) How are the dipole transitions made allowed
since in many instances, the local environment has inversion symmetry, and in that
case\cite{Manceau92}, the transition is symmetry forbidden? There have been
a large number of papers in the last five years addressing these issues for 
many of the transition metals, not all of which are in agreement.
However some  questions have been answered. Since quadrupole-allowed pre-edge features have 
a strong angular dependence, in contrast to the dipole-allowed transitions, 
measurements on single crystals, as a function of angle can separate the two contributions.
Such studies have shown that quadrupole transitions contribute to the A$_i$ peaks in 
Ti,\cite{Joly99,Aifa97} V,\cite{Poumellec98,Sipr99} Ni,\cite{Heumann97} 
and Fe,\cite{Heumann97} with the largest contribution at the lowest energies of the 
pre-edge. The amplitude can be as large as $\sim$ 4\% of the absorption edge height
for some systems at optimum orientations; but more generally it is  
of order 1\%, and could be smaller in powdered samples which are orientational 
averages. The dipole-allowed A$_i$ peaks are often in the 5-15\% range and often
do dominate, but not always. For example, for Ti in rutile
(TiO$_2$), the small A$_1$ peak appears to be primarily a quadrupole feature.\cite{Joly99}

Early Mn XANES work\cite{Manceau92,Belli80} 
assumed that the A$_1$-A$_2$ splitting is produced by the crystal 
field parameter, often called the 10Dq
parameter, which splits the t$_{2g}$ and e$_g$ states. These investigations did 
not consider the possibility of a large on-site Coulomb term, U. Recent work, 
using the Local Spin Density Approximation (LSDA or sometimes LDA) with and 
without U, and including the Hund's rule exchange parameter, J$_H$, find a Coulomb
splitting of both the t$_{2g}$ and e$_g$ states, with the e$_g$ states 
further split by the Jahn-Teller (J-T)interaction.\cite{Satpathy96}

Pickett {\it et al},\cite{Pickett96} (LSDA model) suggest that these
systems are half metallic, with a gap between the O band and a minority spin 
d-band.  They also point out that near 25 \% Ca, all Mn sites could be 
essentially identical if the Ca were uniformly distributed such that there are 
two Ca and six La second neighbors to each Mn. Thus for the concentration range 
20-30\%, the local environment for each Mn may be very similar.
Anisimov {\it et al.} and
Mizokawa {\it et al.}\cite{Anisimov97,Mizokawa97} suggest that a large
fraction of the d-electrons are found on the Mn atoms rather than being
transferred to the O atom as in an ionic solid (thereby leaving holes in
the O band). These calculations yield nearly the same electron density on
each Mn atom, for sites associated with formal Mn$^{+3}$ and Mn$^{+4}$ valences.
Other recent papers\cite{Dai96a,Zhao96,Zhou97,Kwei97c} have also
stressed the importance of O, and the question of charge localization on
the O atoms\cite{Ju97} or on the Mn atoms has been considered.

Some promising calculations for considering the pre-edge features are those
of Elfimov {\it et al.}\cite{Elfimov99} These
calculations indicate that in addition to U and J$_H$,
there are appreciable higher order Coulomb terms that must be included and 
that strong hybridization occurs between the Mn 4p orbitals and the 
Mn 3d states on neighboring Mn atoms. The resulting
splitting of the majority and minority  e$_g$ spin states results in
a splitting of the Mn pre-edge features.
We consider these calculations together with some of the new results
on pre-edges in the discussion section.

In this paper we address the valence question and probe the Mn 3d bands using
the near edge structure. Specifically we show there is no Mn $K$-edge shift 
(within 0.04 eV) through T$_c$. We also compare the main edge, which is too 
narrow to arise from a mixture of ionic Mn$^{+3}$ and Mn$^{+4}$, with the edge
for a material, Sr$_3$Mn$_2$O$_{6.55}$, that does have a mixture of these ions. 
In addition, using a careful subtraction
method, we show that there is indeed a small structure in the main edge that
correlates with T$_c$ for the CMR samples and with T$_{CO}$ for the charge ordered
material. The structure for CMR  material is out of phase with that for the CO 
sample - which suggests that there is a distortion for the CO sample that increases
at low T. We also note that the pre-edge structure has a temperature
dependence which again correlates with T$_c$ for CMR samples. The 
splitting of the pre-edge peaks decreases in the ferromagnetic
phase which may indicate a change in covalency. Finally our interpretation 
of the XANES differs from earlier work on Mn $K$-edges but is consistent
with recent studies of other transition metal atoms.

In Sec II we summarize the samples and experimental setup; some
details were given earlier.\cite{Booth98b} Then in Sec III, we provide a
more extensive discussion of the shift of the Mn $K$-edge as a function of
concentration and temperature. Here we also present the pre-edge results.
We consider the implications of these results in Sec IV.

\section{Experimental details}
\label{exp_details}

Many samples are used in this study, with the average Mn valence changed in
a variety of ways:- divalent substitutions for La$^{+3}$ and
changes in the La$^{+3}$ or O concentrations. 
Powder samples of La$_{1-x}$A$_x$MnO$_3$ were prepared by solid state reaction of
La$_2$O$_3$, MnO$_2$, and a dopant compound - CaCO$_3$, PbO, BaO, for various 
divalent atoms, A.  Ca substitutions are $x$=0.0, 0.12, 0.21, 0.25, 
0.3, 0.65 and 1.0, and Ba and Pb are 0.33. Several firings with repeated 
grindings were carried out using temperatures up to 1400$^\circ$C, with in some 
cases a final slow cool at 1$^\circ$C per minute.  The dc magnetization was 
measured using a commercial SQUID magnetometer. The end compounds, 
CaMnO$_3$ and LaMnO$_{3.006}$,
show antiferromagnetic transitions at $\sim$130 and 125 K, respectively, 
while the $x$=0.65 sample showed features consistent with a charge ordered
(CO) transition at 270 K and an AF transition 
at$\sim$140 K.\cite{Ramirez96,Chen97} Similar measurements on the
substituted manganites indicates that they are all orthorhombic.
The average Mn valence for several Ca substituted samples was also determined by
titration (Sec. \ref{mainedge}).
See Refs. \onlinecite{Booth98a,Booth98b,Snyder96} for further details.

The LaMnO$_{3.006}$ sample was prepared by grinding
stoichiometric amounts of La$_2$O$_3$ (Alfa Aesar Reacton 99.99\%) and
MnO$_2$ (Alfa Aesar Puratronic 99.999\%) in an Al$_2$O$_3$ mortar
and pestle
under acetone until well mixed. The powder sample was formed into a 3/4"
diameter pellet using uniaxially presure (1000lbs), and fired in an
Al$_2$O$_3$ boat under pure oxygen for 12 hours at 1200-1250$^{\circ}$C.
Next the sample was cooled to 800$^{\circ}$C, re-ground, re-pelletized,
and refired at 1200-1250$^{\circ}$C for an additional 24 hours. This process was
repeated until a single phase, rhombohedral XRD trace was obtained.
The reground powder was placed in an
Al$_2$O$_3$ boat and post-annealed in UHP Ar at 1000$^{\circ}$C for 24 hrs. The
oxygen partial pressure was about 60 ppm (determined using an Ametek oxygen
analyzer). The sample was then quenched to room temperature. Diffraction,
titration and TGA measurements
indicate this sample is essentially stoichiometric,
with an oxygen content of 3.006.

Additional LaMnO$_{3+y}$ specimens with various average Mn valences were prepared 
at 1300 $^{\circ}$C in air, followed by three intermediate regrindings.  
The original specimen was removed 
from the furnace at 1300 $^{\circ}$C and has a Mn valence of 3.150.  
A piece of this specimen was reacted 
overnight at 1000 $^{\circ}$C and removed from the furnace, producing a 
sample with an average Mn valence of 3.206.  A nearly stoichiometric
specimen with average Mn valence of 3.063 was prepared at temperatures up
to 1350 $^{\circ}$C with 4 intermediate regrindings in flowing helium gas.
Finally, the nonstoichiometric La$_{0.9}$MnO$_3$ specimen was prepared at 
temperatures of up
to 1350 $^{\circ}$C with three intermediate regrindings.  It was slow-cooled
in air at 1.5 $^{\circ}$C/min to room temperature and had an average Mn
valence of 3.312. For each of these samples the valence was determined
by titration.

A sample that should have isolated Mn$^{+3}$ and Mn$^{+4}$ species is also 
needed for comparison purposes; such a material is 
Sr$_3$Mn$_2$O$_{6.55}$.\cite{Mitchell98} The two species are 
due to the oxygen defect structure 
that puts vacancies into the MnO$_2$ planes to form mixtures of square 
pyramids and octahedra.  This highly insulating material can then be understood 
from chemical reasoning to be Mn$^{+3}$ (square pyramids) and 
Mn$^{+4}$  (octahedra).  Some further justification for this assignment comes 
from the compound Ca$_2$MnO$_{3.5}$, which is all Mn$^{+3}$
and has only square pyramids with vacancies in the MnO$_2$ planes;  it is an
ordered superstructure of the single-layer compound.\cite{Leonowicz85}
Sr$_3$Mn$_2$O$_{6.55}$ was synthesized by firing a stoichiometric mixture of
SrCO$_3$ and MnO$_2$ at 1650 $^\circ$C for 12 hr followed by rapid quenching
into dry ice. This procedure is essential to prevent decomposition into
$\alpha$-Sr$_2$MnO$_4$ and Sr$_4$Mn$_3$O$_{10}$ on cooling and to prevent
oxidation to Sr$_3$Mn$_2$O$_7$.  The oxygen content was measured independently
by iodometric titration and by thermogravimetric analysis, both techniques
yielding  6.55(1) oxygen atoms per formula unit.

All XAFS data were collected at the Stanford Synchrotron Radiation Laboratory
(SSRL).  Most  Mn $K$-edge data were collected on beam line 2-3 using Si(220)
double monochromator crystals for all samples.  Some data were collected on
beam line 4-3 using Si(111) crystals, while most of the Mn $K$-edge
data for the Ba substituted sample were collected on beam line 10-2 using
Si (111).  The manganite powders were reground, passed through a
400-mesh sieve, and brushed onto scotch tape.  Layers of tape were stacked
to obtain absorption lengths $\mu_{\rm Mn}t\sim$1 ($\mu_{\rm Mn}$ is the
Mn contribution to the absorption coefficient and $t$ the sample
thickness) for each sample.  Samples were placed in an Oxford LHe flow
cryostat, and temperatures were regulated to within 0.1 K.
All data were collected in transmission mode. A powdered Mn metal sample
was used as an energy reference for each scan.  The pre-edge absorption
(absorption from other excitations) was removed by fitting the data to a
Victoreen formula, and a simple cubic spline (7 knots at 
constant intervals $\sim$140 eV in $E$) was used to
simulate the embedded-atom absorption, $\mu_0$, above the edge.

The edge shifts are reported relative to a Mn powdered metal foil for which
we take the position of the first inflection point to be
6537.4 eV.\cite{Subias97} For each scan, the position of the
reference edge was determined by fitting the edge to that of a fiducial scan.
This provided a correction to the relative edge position consistent within
$\pm$ 0.015 eV - see next section.

In the pre-edge region there is a remnant of the La L$_{I}$ XAFS that 
must be considered; the oscillation amplitude is about  0.3 \% of the Mn
step height, just before the pre-edge. However, the La $K$-edge XAFS 
show that there is a "beat" 
in the XAFS from about 8.4-10 \AA$^{-1}$, which for the La L$_{I}$ XAFS 
corresponds to the range of the Mn XANES. In this beat region the La L$_{I}$ 
XAFS is reduce by another factor of 4; thus the La oscillations underlying 
the Mn XANES region has an amplitude of about 0.08\%, much smaller than the 
changes we investigate. In addition, this oscillation is slowly
varying with energy, and would at most produce a slowly varying background. 
Consequently any remaining La L$_{I}$ XAFS are not a problem for the Mn 
XANES study.

\section{Near edge results}
\subsection{Main edge}
\label{mainedge}

In Fig. \ref{edges} we show the Mn absorption $K$-edge for several concentrations
of Ca,  33\% Ba and Pb, La$_{0.35}$Pr$_{0.35}$Ca$_{0.3}$MnO$_3$, 
a Sr$_3$Mn$_2$O$_{6.55}$ sample 
that should have a nearly uniform mixture of ionic Mn$^{+3}$ and Mn$^{+4}$, 
some O excess samples, and a La deficient sample.

For the Ca substituted samples several
points are immediately obvious: (1) To first order the main absorption edges 
(ignoring pre-edge structures for now) have almost
the same shape for each dopant concentration and shift nearly rigidly to higher
energy as the concentration is increased, (2) the edges for the manganite 
samples are very sharp, roughly half as wide as the edge for the 
Sr$_3$M$_2$nO$_{6.55}$ sample, 
(3) there is no obvious kink or structure in the sharp edges for the 
substituted (La,Ca) manganite samples 
that would indicate a simple mixture of Mn$^{+3}$ and Mn$^{+4}$ ions,
(4) however, there is a tiny shape change, visible in Fig. \ref{edges}
for samples of different concentration, which shifts the {\it position} of the 
inflection point on the edge relative to the half height position.

%

The data for the La$_{0.35}$Pr$_{0.35}$Ca$_{0.3}$MnO$_3$ sample looks 
very similar to that for La$_{0.7}$Ca$_{0.3}$MnO$_3$
indicating that replacing some of the La by Pr does not change the local 
electronic configuration on the Mn.  The O excess and La deficient samples 
show a similar edge shape to LaMnO$_3$, but the edge shift is considerably 
smaller than expected based on the Mn valence obtained from TGA. The 
shifts for the O excess data are inconsistent with data from other 
groups\cite{Subias98,Maurin99} and are included here to show the 
sharpness of the edge. However, such data suggest that the position of the 
Mn $K$-edge is determined by several factors and using the Mn valence and 
O content obtained from TGA may not be sufficient.

In contrast to the Ca substituted materials, Ba and Pb substitution results in
a significantly broader edge, more comparable to the edges of other Mn 
oxides\cite{Manceau92,Liu97,Yamaguchi98} and the 
Sr$_3$Mn$_2$O$_{6.55}$ sample. There is relatively more weight in the lower part
of the edge.

%

In addition to the shift of the inflection point position on the edge with 
Ca concentration, as noted above, the region of steepest slope 
is also quite broad. Consequently using the position of the peak in the 
first derivative curve as a measure of the average edge position (as we 
and others have done previously) is only an approximate measure of the average 
edge shift. Using the derivative peak yields a roughly linear shift with 
concentration.\cite{Booth98b,Subias97,Croft97} Our data and that of 
Sub\'{i}as {\it et al.}\cite{Subias97} have the same edge shift per valence unit,
while the shift reported by Croft {\it et al.}\cite{Croft97} is smaller. 
This may be the result of different O content in the samples.

To obtain a better estimate of the average edge shift with concentration
(at room temperature), we have fit the LaMnO$_3$ edge data (or the CaMnO$_3$ data) 
to that for each of the other samples, over the main part of the edge (above the 
pre-edge structure).  In this procedure it is 
important that when the absorption from other atoms is removed, the data 
base-line below the pre-edge structure be at zero. Each edge is also
normalized using some feature of the data; for the data at different 
concentrations, we normalized over a range of energies well above the edge,
where the XAFS 
oscillations are small. Similarly we fit the corresponding 
reference edges (Mn foil) to a reference scan to obtain a net overall edge shift.
Several examples of these fits are shown in Fig. \ref{edge_fits}. 
Although there is a change in shape between LaMnO$_3$ and CaMnO$_3$, 
the relative shifts determined with either end compound are nearly
identical - less than 0.02 eV difference over the entire concentration range.


In Fig. \ref{valence}a we plot the relative shifts obtained 
from fits to LaMnO$_3$ at room temperature. The shift with 
x is roughly linear with concentration, with a net shift from 0 to 100\% Ca of 
$\sim$3 eV. This is considerably smaller than the value 4.2 obtained from the 
derivative peak\cite{Booth98b,Subias97} and illustrates the effect of the 
shift of the inflection point relative to the half height. However, over the 
straight part of the plot from x= 0.3-1.0, the slope is 3.3 eV/valence unit,
quite close to the 3.5 eV/valence unit obtained by 
Ressler {\it et al.}\cite{Ressler98} for MnO, Mn$_2$O$_3$, and MnO$_2$.  The 
point at x = 0.12 is anomalous, but titration measurements
give about the same Mn valence for the 12 and 21\% samples, which agrees 
with the comparable edge shifts.  The same data is re-plotted as a function
of the titrated valence in Fig. \ref{valence}b; in this
case the variation with valence is smoother, but slightly non-linear. The
different values for the titrated valence, compared to the value expected
from the Ca concentration may indicate that there are slight
variations in O content in some samples.


We have also used  a similar analysis to investigate any possible edge shift 
as a function of temperature, by fitting the entire edge of the 50K data  for 
a given sample to all the higher temperature data files. Two examples are shown in 
Fig. \ref{edge_fits}c,d.  This figure shows that changes in the shape of the 
main edge, above and below T$_c$, are quite small (although measurable). The 
largest relative change is in the pre-edge peaks, to be discussed later. 
In Fig. \ref{edges_temp} we show the shift of the edge position as a 
function of temperature up to 320K for several sets of samples. 
(We have similar data out to nearly 500K for the 
Ba sample and CaMnO$_3$.)  Variations in the fit values for the net
shift, $\Delta$E$_o$, for several traces at the same temperature
are less than $\pm$ 0.02 eV, and fluctuations about the small average 
shift with T are comparable for a given experimental run. Differences
between experimental set-ups or using Si$<$111$>$ or Si$<$220$>$ monochromators
are less than 0.1 eV.
For T less than 300K, the net shift for each sample is very small 
(less than 0.04 eV), but nearly all appear to have a slight decrease at 
high T.

\subsection{Pre-edge region}

In Fig. \ref{pre-edge} we plot the pre-edge region as a 
function of temperature on an expanded scale for CaMnO$_3$, a CMR sample 
with 21\% Ca, and the 33\% Ba sample. Data for LaMnO$_3$, the CO sample 
with 65\% Ca and another CMR sample have recently been published\cite{Bridges00}
in a short paper. For these systems the 
main features are the lower three peaks labeled A$_1$-A$_3$ (near 6539, 
6541, and 6544 eV) and the B peak. The lower two peaks A$_1$ and A$_2$ are
common to all materials although not resolved for the 33\% Ba data 
collected using Si (111) crystals which have a lower energy resolution. 
The comparison of the two Ba data sets in this figure 
illustrate the importance of using high energy resolution. The A$_{3}$ peak
is not obviously present in most samples. 
In Fig. \ref{pre-edge3}
we compare the data for the 30\% Ca CMR sample with the Sr$_3$Mn$_2$O$_{6.55}$
sample and also show the pre-edge for the Pb sample, all on a more expanded 
scale.

There are several features to note;  all the pre-edge 
features start at very nearly the same position regardless of doping and 
the amplitude of the pre-edge features labeled A increases with average 
Mn valence (Ca 
concentration) as observed in other Mn compounds\cite{Manceau92} and in a 
previous manganite study.\cite{Croft97}  There are, however, small shifts of 
these features with Ca concentration as shown in Fig. \ref{pre-edge_shift}.  
The A$_1$ peak energy increases slightly from LaMnO$_3$ to 
CaMnO$_3$, and the A$_1$-A$_2$ splitting decreases from 2.2 to 1.8 eV. 
(The exception is the 65\% sample, but here the A$_i$ peaks are poorly resolved.)
For the substituted samples, the leading edge of the A$_1$ peak remains 
steep for all concentrations except the 65 \% sample. Consequently, the 
pre-edge for the intermediate concentrations (CMR samples) cannot be 
modeled as a simple weighted sum of the end compounds LaMnO$_3$ and CaMnO$_3$.
Note that the leading edge for the Sr$_3$Mn$_2$O$_{6.55}$ 
sample (See Fig. \ref{pre-edge3}) is broader, consistent with a mixture of 
Mn$^{+3}$ and Mn$^{+4}$ ions, and also has a significant A$_3$ peak. 
The latter is not present in the data for 30\% Ca.

The Ba and Pb pre-edges are slightly different; the Ba pre-edge features are 
not as well resolved even for the higher energy resolution data while 
the A$_1$ peak is largest for the Pb sample (Compare Fig. \ref{pre-edge3}
with Fig. \ref{pre-edge}). 


The most striking feature in Figs. \ref{pre-edge},\ref{pre-edge3} is the 
variation in the 
intensity of the pre-edge peaks and the shift of A$_2$ as T increases through 
T$_c$ for the CMR samples. In contrast, the change for LaMnO$_3$ is small 
up to 300K.\cite{Bridges00} For the 21\% Ca sample in Fig. \ref{pre-edge}
the A$_1$ peak decreases in amplitude while the A$_2$ and B-peaks increase 
with increasing T; the A$_2$ peak is sharpest at 300K and clearly shifts 
downward below T$_c$ (0.4 - 0.5 eV, depending on the background function 
used). See the solid triangles in Fig. \ref{pre-edge_shift}.  The change 
in the A-peaks for the 33\% Ba sample (using the high resolution monochromator)
appear to follow the same trend as observed for the Ca data 
(Fig. \ref{pre-edge3}) but the A$_2$ peak is not as well resolved. 


The largest temperature dependence is observed for the 
CaMnO$_3$ sample above 300K (See Fig. \ref{pre-edge}a), with the largest 
increase occurring for the B-peak. Also, the amplitude of the peak at the 
top of the edge, commonly called the ``white line'', (see Fig \ref{edges}c 
at 6554 eV for example) decreases slightly at high T.
These effects become much larger at only slightly higher temperatures and
will be treated in a separate paper.  For the CMR samples, we associate 
the temperature dependent changes in the amplitude of the pre-edge features 
with changes in charge localization/hybridization.

\subsection{Difference Spectra}
More detailed information can be obtained by examining the change in the shape 
of the XANES region as a function of temperature. The files are first shifted to
correct for any small changes in the energy of the monochromator 
and all spectra are carefully normalized as 
discussed earlier.  The difference spectra 
are obtained by subtracting the data at 300K from all the 
data files (at different temperatures) for a given 
sample. This approach was used originally to investigate the pre-edge 
region for the 21\% sample,\cite{Bridges00} but considerable structure 
was found at energies corresponding to the main edge, for both the CMR 
and CO (x=0.65) samples. Several examples of these difference spectra are 
shown in 
Fig. \ref{diff_spec} for LaMnO$_3$, CaMnO$_3$, and the 21, 30, and 65 \% 
Ca substituted samples. 

%

In the pre-edge region, the temperature variation of the A$_1$ and A$_2$ peaks for 
the 21 and 30\% Ca (CMR) samples is very clearly visible in Fig. \ref{diff_spec}c,d;
it begins at T$_c$, 
with most of the change occurring over a 60K range just below T$_c$. 
The temperature-dependent changes of the pre-edge are comparable in both 
samples, with the magnitude of the change 
of the A$_2$ peak being roughly 50-70\% that of the A$_1$ peak. 
For the 65\% Ca sample,
changes of the A$_i$ with T are also observed in the different spectra, but 
the amplitudes are considerably smaller, and interestingly, the 
phase is inverted - the A$_1$ difference peak decreases instead of increasing. 
For the  LaMnO$_3$ sample (Fig. \ref{diff_spec}a) there is essentially no 
structure in the difference 
spectra over the pre-edge energy range, but surprisingly 
there are small peaks in this range for CaMnO$_3$ (See lower part of 
Fig. \ref{diff_spec}a),
with the largest peak in the difference spectra occurring {\it between}
A$_2$ and A$_3$ - this suggests
that there 
\noindent are in reality more than three pre-edge peaks.

There is also well defined structure in the difference spectra over the 
energy range of the main edge, although it is only a few percent of the 
edge in amplitude. For LaMnO$_3$ there is a broad feature over most of the 
edge region which increases as T is lowered. CaMnO$_3$ has a similar feature
but it is larger and narrower 
(Fig. \ref{diff_spec}a). Both appear to correspond to the temperature 
dependent peak near the top of the edge (the ``white line'' mentioned
earlier) which is sharpest at low T. For the CMR samples, there is additional 
structure on top of this broad peak - a dip at 6551 and a peak at 6553 eV 
(2 eV apart). 
Another dip/peak occurs just above the edge at 6555-6556.5 eV. 
The CO sample also shows structure over this energy range but again the phase 
is inverted relative to the CMR samples (i.e. a peak/dip at 6552 and6554 eV),
- this phase inversion thus extends over the entire near-edge region. 

\section{Discussion}

\subsection{Main edge}

For Mn atoms the main $K$-absorption edge represents transitions mainly from 
the atomic 1s state to the empty Mn 4p band.
The XANES results show that this edge is very sharp for the Ca-substituted
samples, the O-excess samples and the La deficient sample. The width of
the edge (roughly 5-6 eV) is narrower than the edge for most other
Mn compounds and the shift in edge position is $\sim$ 3 eV for a valence
change of +1. No obvious indication of a step or double edge structure is
present that would indicate two distinct valence states.
If completely localized Mn$^{+3}$ and Mn$^{+4}$ ions were present on time
scales of 10$^{-14}$ sec, the edge should have a smaller average slope and
generally be broader, as would be expected for a mixture of fine powder
of LaMnO$_3$ and CaMnO$_3$. To model this explicitly, we compare in
Fig. \ref{sim_65} the experimental edge for the 65\% Ca (CO) sample and a
weighted sum of the +3 and +4 end
compounds; clearly the experimental edge is much sharper as noted
previously.\cite{Booth98b,Subias97}


In contrast to the Ca-doped samples, the edge for Sr$_3$Mn$_2$O$_{6.55}$ is
much broader (Fig. \ref{edges}a), with a width of 10-12 eV. There is also
a change of slope of the main edge for this sample that is consistent with
two valence states, but the shape is more complicated.
Note that a combination of two edges each $\sim$5 eV wide, separated by
$\sim$3 eV (the
separation for a valence change of 1) would yield an
edge of width $\sim$ 13eV. Thus the width and structure of the
Sr$_3$Mn$_2$O$_{6.55}$ edge are both consistent with the expectation
that two valence states are present in this sample --
Mn$^{+3}$ and Mn$^{+4}$.  Similarly our data for Mn$_3$O$_4$, which
has a mixture of +2 and +3 valence states, has a very broad edge of
13-14 eV (not shown).

The edges for the Ba and Pb samples are also broader; they have
a break in slope and more amplitude in the lower part of the edge that
might suggest two valence states. In this regard they are quite
different from the Ca substituted samples for which the shape of the
main edge for the CMR samples does not change much from that
of LaMnO$_3$. In addition, the net shifts of the edges for the 33\% Ba,
and Pb samples are smaller than expected; the additional structure
near 6546-48 eV shifts the lower part of the edge down in energy while the
top of the edge is close to the position of the 30\% Ca samples. The net
result is a very small overall average edge-shift compared to LaMnO$_3$. 

We have also shown by fitting over most of the edge, that there is 
no significant change in the average edge position for any of 
the substituted manganite samples near T$_c$. This agrees with
our earlier result\cite{Booth98b} in which we averaged data points above
and below T$_c$. The new analysis also indicates that there is consistently
a slight decrease in edge position at the highest temperatures, that
is largest for CaMnO$_3$ and Sr$_3$Mn$_2$O$_{6.55}$.  The reason for this 
downward shift is not yet clear but may be related to the temperature
dependence of the B peak (since an increase in the B-peak intensity effectively
shifts the lower section of the edge to lower energy).

However, the lack of any temperature dependence below T$_c$ disagrees with 
the earlier work of 
Sub\'{i}as {\it et al.},\cite{Subias97} who report a 0.1 eV decrease 
in edge position up to T$_c$ for La$_{0.6}$Y$_{0.07}$Ca$_{0.33}$MnO$_3$ 
and then a 0.09 eV increase up to 210K. The small but important change
in the shape of the main edge in Fig. \ref{diff_spec} provides a partial
explanation for this discrepancy.
Sub\'{i}as {\it et al.} assumed no change in edge shape and calculated difference
spectra  for each temperature. Under this assumption, the amplitude of the
peak in the difference spectra would be proportional to the energy shift of the
edge. The additional structure observed in the difference spectra indicates 
there is a shape change rather than an overall edge shift.

The difference spectra (Fig. \ref{diff_spec}) also show clearly that the
additional structure is dependent on T$_c$. For the CMR samples, the 
dip-peak structure, superimposed on the peak observed for the end compounds,
begins to be observable near T$_c$ and grows rapidly in the 60-100K
range just below T$_c$. This structure means that compared to the edge at 
low temperatures (undistorted Mn-O bonds)
the edge above T$_c$ (distorted Mn-O bonds)
has the upper part of the edge shifted upward in energy
while the lower part is shifted downwards. The separation between 
the dip-peak structure is about 2 eV (see vertical dotted lines). 

This raises several questions - for the CMR samples is there a mixture of
+3 and +4 sites as usually assumed? If so why is the edge structure so 
small? Can the 
small structure observed in the difference spectra  be explained in 
some other way? One aspect that must be included is the very large width 
of the Mn 4p density of states (DOS), roughly 15 eV wide, that is found 
in two quite different 
recent calculations.\cite{Elfimov99,Benfatto99} The main edge is due to
transitions into this band and the calculated absorption
edge\cite{Elfimov99} (broadened by the core-hole lifetime) is very 
similar to that observed experimentally.\cite{Bridges00}
The broad width means that the 4p states are
extended and not localized on one Mn atom. Consequently,
the $K$-edge will correspond to a Mn valence partially averaged 
over several Mn atoms, and thus will be less sensitive to variations 
in local charge on different Mn sites.

Another possibility is that the system is more covalent and that there
are some partial holes in the O 2p band which is hybridized with the Mn 3d
states. This is supported by several calculations and by the observation
of holes in the O 2p band in absorption studies\cite{Ju97}. Such holes may 
play an important role in the unusual transport of these materials. In
calculations, Anisimov {\it et al.}\cite{Anisimov97} and 
Mizokawa and Fujimori\cite{Mizokawa97}
obtain two types of Mn e$_g$ configurations with almost identical local 
charge densities. In both calculations
there are distortions of the Mn-O bond distances. For the calculation
of Anisimov {\it et al.}\cite{Anisimov97}, one configuration 
is symmetric in the $ab$ plane with four small equal lobes directed towards
O while the other 
has two large (and two small) lobes, again directed towards O atoms in the
$ab$ plane. The more symmetric case is
associated with a  formal Mn$^{+4}$ site and the other state with Mn$^{+3}$
but because the charge densities are comparable would not lead to 
significantly different edge shifts.
The recent calculations of Elfimov {\it et al.}\cite{Elfimov99}
are also relevant. To fit the observed splittings of A$_i$ peaks (2.2 eV for 
LaMnO$_3$ and 1.8 eV for CaMnO$_3$), U and J$_H$ had to be lowered from the 
values in the first calculation\cite{Elfimov99} - to 4 eV and 0.7 eV, 
respectively,\cite{Bridges00} which implies higher covalency.

The remaining question to be answered about the main edge for the CMR samples
is what is the explanation for the small structure in the difference spectra
as T is lowered below T$_c$?  A possible answer is again found in the calculations
of Elfimov {\it et al.}\cite{Elfimov99}. They find that the position of the 
4p partial DOS is bond-length dependent - it occurs at a lower energy
when the Mn-O bond lengthens (p$_x$ orbitals in their paper) and is at
a higher energy for shorter bond lengths ((p$_y$ and p$_z$).
Such a shift is expected; in polarized XAFS experiments on high T$_c$ materials
we have observed edge shifts between the c- and a-axes. In addition, studies
of molecules show that the edge shifts to higher energy when the bond length
shortens.\cite{Bianconi88}
The separation between the partial DOS for p$_x$ and p$_y$ is about 2 eV
in Elfinov {\it etal}'s calculation when they use distortions similar to those 
observed in LaMnO$_3$; we expect to see some evidence of this splitting 
in the experimental absorption edge, although it is lifetime broadened
and the 4p states are extended. We propose that the tiny dip-peak
structure observed for the CMR materials is the result of  the 
different positions for the partial DOS for p$_x$ and p$_y$. The
dip-peak splitting is also about 2 eV but it is not clear whether this 
is significant or a coincidence.

For the CO sample there is also structure in the edge but the phase is inverted.
If the above explanation for the {\it dip-peak} structure in the difference spectra
for CMR samples is correct then it suggests that the {\it peak-dip} feature for the
CO sample is also produced by local distortions - but in this case by a
local distortion that starts at the charge ordering temperature, T$_{CO}$=270K, 
and {\it increases} as
T is lowered. Such a model then provides a simple interpretation 
for the unusual lack of temperature dependence (reported\cite{Booth98b} but 
not explained) for 
$\sigma^2$ for this sample. The surprise is that at least the thermal phonon 
broadening should have caused some increase in $\sigma^2$ with T.
However, if there is a distortion associated with the CO state, then 
there must be an associated broadening contribution $\sigma_{CO}$, for the 
Mn-O pair distribution
function, $\sigma^2_{CO}$, that is zero above T$_{CO}$ and increases as T 
is lowered below T$_{CO}$. Then the 
total variance for the Mn-O bond, $\sigma^2_{Mn-O}$, will be given by 

\begin{equation}
\sigma^2_{Mn-O}(T) = \sigma^2_{phonon}(T) + \sigma^2_{CO}(T) + \sigma^2_{static}
\end{equation}

\noindent where $\sigma^2_{phonon}(T)$ is the phonon contribution,
and $\sigma^2_{static}$ is a static (temperature-independent) contribution
from disorder. $\sigma^2_{phonon}(T)$ should be comparable to that for
CaMnO$_3$, since we see the same phonon component for both pure CaMnO$_3$
and La$_{0.79}$Ca$_{0.21}$MnO$_3$  above T$_c$. To make $\sigma^2_{total}$
nearly independent of T means $\sigma^2_{CO}(T)$ and $\sigma^2_{phonon}(T)$
almost cancel for temperatures below 300K.


We can extract the CO contribution
following the method in Ref. \onlinecite{Booth98b} for calculating $\Delta \sigma^2$
- we fit the two highest T data points (from Ref. \onlinecite{Booth98b}) to 
$\sigma^2_{phonon}(T)$ + $\sigma^2_{static}$
and then subtract these contributions from the data. In Fig. \ref{D_s2_CO} we
plot the result of this analysis for the 65\% Ca sample. From
this figure, the maximum value for $\sigma^2_{CO}(T)$, is roughly
10\% of  that associated with polaron formation for the
CMR samples. Such an increased distortion for charge ordered material makes sense
- as the sample
becomes charge or orbital ordered, there is more room for the longer Mn-O bonds
to lengthen, while for a random arrangement of orbitals, the series of long
and short Mn-O bonds are more constrained.

\subsection{Pre-edge region}
\label{preedge_region}

\subsubsection{Background}

The pre-edge results provide additional information about the nature of the
electronic states. For many of the transition elements, 1-3
pre-edge peaks, A$_i$,  occur well below the main edge ($\sim$ 15eV below) and are 
assigned to transitions to empty states with d-like character,
i.e. these are 1s to 3d$^{(n+1)}$
transitions\cite{Griffith61,Sugano70} where  n is the initial number of d electrons 
and  n+1 includes the excited electron in the final state, which usually includes 
the effect of a core hole. The 1s-3d transitions are directly allowed through the very weak
quadrupole transition\cite{Manceau92,Joly99,Aifa97,Poumellec98,Sipr99,Heumann97} or allowed 
via an admixture of 3d and 4p states\cite{Manceau92,Joly99,Poumellec98,Heumann97,Bridges00}. 
In the pre-edge region there may also be some hybridization with the O 2p states.
If the metal site is centrosymmetric, there is no mixing of 3d and 
4p states on the excited atom and 1s-3d dipole transitions are strictly 
forbidden;\cite{Manceau92} however local distortions
can make such 1s-3d transitions very weakly dipole allowed. Three aspects need to be 
recognized in considering the Mn pre-edge in the present study.

\begin{itemize}

\item {\it Quadrupole interactions}\\
Although the quadrupole interaction is weak it has a clear signature through the 
angular dependence of the absorption process. Based on recent studies on oriented 
single crystals\cite{Joly99,Aifa97,Poumellec98,Sipr99,Heumann97}
we estimate that quadrupole-allowed peaks will be at most 1\% in powder samples 
which is considerably smaller than the A$_i$ peaks observed for the substituted manganites
but perhaps not negligible. A small quadrupole component, as seen for 
example in FeO,\cite{Heumann97} may well be present. 

\item {\it Dipole allowed via 3d-4p mixing on the absorbing atom}\\
If the Mn site lacks inversion symmetry, then in principle there will be mixing of the
Mn 3d and 4p states on the central atom. Consider a system that is nearly cubic
but has a small distortion that removes the inversion symmetry - i.e. the metal atom is 
slightly displaced such that the bonds on opposite sides of the metal atom are slightly
different. Then the mixing parameter is $\sim$ $\delta_l/r_o$ and the matrix element will
be proportional to $(\delta_l/r_o)^2$, where $\delta_l$ is the difference in opposite
bond lengths and $r_o$ is the average bond length. An example is the V site in V$_2$O$_5$; 
here the VO$_6$ octahedron is strongly distorted,\cite{Poumellec98,Sipr99} and along 
the c-axis, the two 
V-O bond lengths are 1.577 and 2.791 \AA~respectively.  Experimentally there is a large 
pre-edge peak for the V $K$-edge that can be modeled assuming 3d-4p mixing on 
the absorbing atom plus the effect of a corehole. 

\item {\it Dipole allowed via 4p mixing with neighboring metal atom 3d states}\\
In several cases, a significant pre-edge peak is observed in a cubic crystal, which 
can't be explained by the above 3d-4p mixing on the excited atom. Multi-scattering 
calculations for such systems often show that large clusters are
needed before the pre-edge features are produced - scattering paths are needed which 
include many further neighbors, particularly the second neighbor metal atoms. 
An equivalent result emerges from band theory calculations where the hybridization of 
extended states is important. Here the dipole transition can be made allowed via mixing
of the 4p state on the central atom with the 3d states on neighboring atoms. Projections
of the density of states with p character (p-DOS) for such systems show small features 
at the energies of the 3d states; such features are not observed in the p-DOS when the
pre-edge feature is a quadrupole transition. In the limit of multi-scattering 
calculations with very large clusters, the two approaches (band theory and 
multi-scattering) should be equivalent.

\end{itemize}

To have a mixing of 3d with the 4p states (to make a state of p-character), 
one needs a combination
of 3d states that has odd parity as pointed out by Elfimov {\it et al.}\cite{Elfimov99}.
It is easy to obtain such a state if a linear combination of
3d states on two neighboring Mn atoms is used and the p-states are extended enough to
partially overlap them.
Specifically, consider 3 Mn atoms in a line - a central excited atom (0) and
left (L) and right (R) atoms - with $\Psi_{4p}$(0) being the 4p state on the
central atom, and $\Psi_{3d;x^2-y^2}$(R$_R$) and $\Psi_{3d;x^2-y^2}$(R$_L$)
being the 3d$_{x^2-y^2}$ states centered on the right and left atoms. Then a state
with odd symmetry about the central atom is given by

\begin{equation}
\Psi_{total} = \alpha \Psi_{4p}(0) +
\frac {\beta} {\sqrt{2}}(\Psi_{3d;x^2-y^2}(R_L) - \Psi_{3d;x^2-y^2}(R_R))
\end{equation}

\noindent where $\alpha$ is essentially 1.0 and we ignore the intervening O atom
via which the hybridization occurs. The small parameter, $\beta$, is a measure
of the hybridization and is strongly dependent on the overlap of the 4p and 3d
wavefunctions on different Mn atoms, and hence on the distance between them.

\subsubsection{Application to the substituted manganites}

The pre-edge for the substituted manganites follows the general trends observed
for other Mn systems quite well. Three A-peaks are observed for CaMnO$_3$; A$_2$ is
 larger than the A$_1$ peak, and the A$_2$-A$_1$
splitting is smaller (high valence - +4) than for other samples. The LaMnO$_3$ case
is similar; the A$_2$-A$_1$ splitting is largest (lower valence +3) and the overall
A peak amplitude is smallest. However, the A$_2$ peak is larger than expected
from the literature for Mn$^{+3}$ states in other compounds\cite{Manceau92},
possibly because of increased local distortions in this compound. 

However, there are difficulties with some of the earlier 
interpretations\cite{Belli80,Manceau92} in which the dipole allowed transitions
are assumed to originate from a 3d-4p mixing on the excited atom. First
the A$_i$ peaks appear for both distorted and undistorted systems. Second, the
amplitude (particularly for the relatively undistorted system CaMnO$_3$) is too large
to be a 1s-3d transition made allowed by a slight breaking of inversion
symmetry about the excited Mn atom.  Recently, based on the calculations of 
Elfimov {\it et al.}\cite{Elfimov99}, we have interpreted A$_1$ and A$_2$
as dipole allowed via a mixing of Mn 4p states with Mn 3d states on neighboring
metal atoms\cite{Bridges00}. The projected p-DOS in the calculations of
Elfimov {\it et al.}\cite{Elfimov99}
show two features in the pre-edge region, which indicates that dipole-allowed 
transitions should be present. In addition, the broad Mn 4p band obtained
in that work and by by Benfatto {\it et al.}\cite{Benfatto99}, also implies that 
the 4p states are indeed extended - a necessary requirement
for mixing with the 3d states on the neighboring metal atoms.
Similar interpretations have been given recently for other transition metal systems
that are cubic or very nearly so; Fe in FeO\cite{Heumann97} and  Ti in 
rutile\cite{Joly99,Aifa97}.
A mixing with the 3d states on neighboring Ti atoms was also reported in the layered
disulfide TiS$_2$\cite{Wu97}.

The calculations of Elfimov {\it et al.} also show that there is a splitting
of the unfilled 3d bands - the lowest is the majority e$_g$ band (which may be
partially filled via doping); the next two are the minority e$_g$ and t$_{2g}$
bands which partially overlap. The coupling with the t$_{2g}$ is expected to be
smaller since these orbitals are of the form d$_{xy}$, which has reduced
overlap with the Mn 4p in a $\pi$ bonding configuration. The splitting of
these e$_g$ bands depends both on
J$_H$ and on the degree of covalency/hybridization. As reported
recently,\cite{Bridges00} adjusting the parameters in this calculation so that
the theoretical splitting is close to the 2 eV observed experimentally,
resulted in U=4 eV and J$_H$=0.7 eV. These lower values also suggest an
increase in covalency and hence that the charge is shared between Mn and O.
Consequently, there is a non-zero density of holes in the O bands, in
agreement with Ju {\it et al.}\cite{Ju97} and these O-holes may play an
important role in the unusual transport of these materials.
For the CMR samples, the additional decrease in the A$_1$-A$_2$ splitting 
for T$<$T$_c$ may suggest a further increase in covalency.

Finally the temperature dependence of the A$_i$ peak amplitudes is still not 
explained.
Comparing the pre-edges of distorted LaMnO$_3$\cite{Bridges00} with almost
undistorted CaMnO$_3$ (See Fig. \ref{pre-edge}) would suggest that as the 
CMR samples change from distorted above T$_c$ to ordered at low T, the A$_2$
peak would increase relative to A$_1$. Experimentally the reverse is true.
However, we still suggest that the observed temperature dependence arises
from the change in local structure, based on the fact that the changes
for the CMR and CO samples are out of phase for both the pre-edge features
and the structure in the main edge.

\subsubsection{Other Models}
Another general feature that emerges from our data is that although
the main changes occur just below T$_c$, there is also a gradual change
to the fully ordered state as the sample is cooled well below T$_c$, and
the local structure continues to change down to 50K and below. Consequently
there may be clusters formed at T$_c$ that grow as T is lowered.  We have
interpreted our local distortion results earlier in terms of a two component
model.\cite{Booth98a} Within that model, one of these components (fluids) would
correspond to delocalized states - these could be either delocalized holes
or delocalized electrons. 
We also point out that the decreasing distortions observed in EXAFS as T is
decreased below T$_c$ and the corresponding increase in resistivity suggest
a changing average mobility of the charge carriers. Within the model we have
suggested, the fraction of delocalized carriers would increase as T is
lowered.\cite{Booth98b} However, one of these components might also correspond to
the Mn atoms in a cluster, the positions of which are
dominated by small variations in dopant concentration or O vacancies,
possibly leading to a regime with phase separation. Such inhomogeneities
likely  play an import role in these materials. In addition,
Jaime {\it et al.}\cite{Jaime99} have successfully modeled their resistivity
and thermoelectric measurements using a two component system of localized and
itinerant carriers.  The recent calculations
using the Kondo model\cite{Yunoki98a,Yunoki98,Dagotto98} also stress phase
separation but it is not clear how to compare with their results.

\section{Conclusions}
We have addressed several issues related to the Mn valence in the substituted
LaMnO$_3$ materials. Although discussions of these systems often assume
isolated Mn$^{+3}$ and Mn$^{+4}$ states, we observe no change ($<$ 0.02 eV)
in the average edge position through the ferro-magnetic transition for the 
CMR systems (Ca, Ba or Pb doped), and in all cases the total edge shift from 
0 to 300K is $<$ 0.04 eV. Although there is no obvious
step or kink in the edge, expected for two well-defined valence states, there
is a very small shape change that can be observed by taking the difference
of data files at different temperatures. A dip/peak structure develops as T
drops below T$_c$ for the CMR samples; the dip/peak separation is $\sim$ 2 eV
and is consistent with the splitting calculated\cite{Elfimov99} for the
p$_x$ and p$_y$ partial DOS when the 
manganite structure 
changes from an undistorted to a distorted (LaMnO$_3$) lattice. For CMR samples,
such changes in the local distortions below T$_c$ were deduced earlier 
from EXAFS data\cite{Booth96,Booth98a,Booth98b,Subias97}. At low T the CMR
samples are very well ordered, but as T increases there is a rapid increase
in the local distortions up to T=T$_c$; above T$_c$ the change in disorder 
changes slowly. The rapid change just below T$_c$ has been associated with 
the formation of 
polarons. These distortions, now observed in both the XANES and EXAFS data,
indicate some change in the local charge distribution. However the small size 
of the effect in the XANES spectra needs to be understood. In part it 
can be attributed to the extended nature of the broad Mn 4p band which 
tends to make the 1s-4p edge transition an average over several Mn atoms. 
However a change in covalency - specifically a transfer of charge between 
Mn 3d and O 2p states - might also be associated with this structural
change, but produce little change in the edge. Support for this possibility
is obtained from the pre-edge results, summarized below. 

For the CO sample we observe a similar behavior, but in this case the
structure in the difference spectra are inverted relative to that 
for the CMR sample. This indicates that the local distortions increase in 
the CO state below T$_{CO}$.

The pre-edge structure provides additional information about the 3d-bands in
these materials. Two or three peaks are observed, labeled A$_1$-A$_3$.
A$_2$ at 300K is essentially independent of concentration while A$_1$ increases
slowly with x; A$_3$ is only observed for high Ca concentrations.
Following the work of Elfimov {\it et al.}\cite{Elfimov99} we attribute
these peaks to a hybridization of Mn 4p on the excited atom with an ungerade
combination of 3d states on neighboring Mn atoms, i.e. they are not the result 
of splittings of atomic multiplets on the excited atom as is often assumed.
Similar explanations for the pre-edge region have recently been proposed
for several other transition metal $K$-edges.  Consequently the splittings
observed are essentially unaffected by the presence of the core hole and should
be a good measure of the splittings of the e$_g$ bands which are influenced
by the hybridization of the Mn 3d and O 2p states.  This interpretation
of the pre-edge does not depend on small distortions of the crystal and
therefore also provides a simple explanation for the large pre-edge features
observed in the more ordered CaMnO$_3$ material. 

In the calculations of Elfimov{\it et al.}\cite{Elfimov99}, the two 
lowest empty bands are the majority and minority spin e$_g$ bands; the 
minority spin t$_{2g}$ band overlaps the latter but is expected to be more
weakly coupled. U and J must be reduced slightly to
fit the experimental splitting (2 eV) of the A$_1$ and A$_2$ peaks; U=4eV 
and J$_H$=0.7eV. This indicates an increase in the covalency. The additional
small decrease in the A$_1$-A$_2$ splitting below T$_c$ may suggest
a further change in covalency or hybridization.

Thus the picture that emerges is that there is considerable hybridization
of the energy states (Mn 4p and 3d, and O 2p), with
some hole density in the O bands and possibly only small differences in
the charge localized on Mn atoms which have different types of 
e$_g$ orbitals. The possibility of distinct types of orbitals can lead 
to orbital ordering\cite{Anisimov97}, with displacements of the O atoms
forming J-T-like Mn-O bond distortions when the hopping charge is 
localized for times
of order the optical phonon periods. As a result, the possibility that part
of the transport takes place via hole density in the O bands needs to be
considered.  Note that slowly hopping holes on the O sites would lead to 
distorted Mn-O bonds while rapid hopping (faster than phonons) would leave 
the O atom at an average undistorted position.


\acknowledgements
The authors wish to thank G. Brown, C. Brouder, D. Dessau, T. Geballe, J. 
Rehr, G. Sawatzky, and T. Tyson for useful discussions and comments. 
FB thanks K. Terakura for sending some of their unpublished
results.  The experiments were performed at the Stanford Synchrotron Radiation
Laboratory, which is operated by the U.S. Department of Energy, Division
of Chemical Sciences, and by the NIH, Biomedical Resource Technology Program,
Division of Research Resources.  Some experiments were carried out on
UC/National Laboratories PRT beam time.  The work is supported in part by NSF
grant DMR-97-05117.


\newpage


\begin{figure}
\caption{The Mn $K$-edges for a number of similar manganite systems, corrected
for energy shifts of the monochromator (see text).
}
\label{edges}
\end{figure}


\begin{figure}
\caption{Fits of absorption edges to each other. In a) and b) we show fits of
the LaMnO$_3$ edge (solid curve) to the 25\% and 100\% Ca samples. In c) and d)
we show fits of the 50K data (solid curve) to high temperature data for two
different samples. Corrections from the reference Mn foil have not been
included in this figure.
}
\label{edge_fits}
\end{figure}

\begin{figure}
\caption{The shift of the Mn $K$-edge with a) Ca concentration and b)
titrated valence for La$_{1-x}$Ca$_x$MnO$_3$. Note we do not have titration
data for x=0.30.
The relative errors are: edge shift, $\pm$ 0.02 eV, smaller than symbols; Mn
valence $\pm$ 0.02 units, slightly larger than symbols. In this and several
following figures, the lines are a guide to the eye only.}
\label{valence}
\end{figure}

\begin{figure}
\caption{The shift of the absorption edge for each sample relative to the low
temperature data. In panels a)-c), one of the curves (squares) is shifted
downward by 0.1 eV for clarity.  Relative errors $\pm$ 0.015 eV; $\pm$ 0.03 eV
for LaMnO$_{3.03}$.
}
\label{edges_temp}
\end{figure}

\begin{figure}
\caption{The temperature dependence of the pre-edge region for CaMnO$_3$,
21\% Ca, and 33\% Ba; note different temperature ranges.
The CaMnO$_3$ sample shows little change of the pre-edge region below 300K
while the A$_1$ and A$_2$ peaks are temperature dependent
for the CMR samples. The Ba pre-edge
data, collected using Si(220), are much sharper than data collected using
Si(111) crystals, and show a splitting of the A peak. Note that the data
collected using Si(220) in c) have been shifted down by 0.15 eV for clarity.
}
\label{pre-edge}
\end{figure}

\begin{figure}
\caption{The temperature dependence of the pre-edge region for
La$_{0.7}$Ca$_{0.3}$MnO$_3$, Sr$_3$Mn$_2$O$_{6.55}$  and the 33\% Pb
samples up to roughly 300K. The leading edge of the A$_1$
peak is broader for Sr$_3$Mn$_2$O$_{6.55}$ than for the 30\% Ca sample in a).
In b) the Pb samples has the largest A$_1$ peak at low T.  Note that the
scale is expanded compared to the previous figures and the B peak is not shown.
}
\label{pre-edge3}
\end{figure}

\begin{figure}
\caption{The positions of the pre-edge features as a function of Ca
concentration - open symbols, and solid circles, 300K.  The error in peak position is
$\pm$ 0.1 eV. For the CMR samples, the A$_2$ peak is shifted to a
lower energy for T $<$ T$_c$ - the filled triangles are for 50K.
}
\label{pre-edge_shift}
\end{figure}

\newpage
\begin{figure*}
\caption{The difference spectra as a function of temperature for 21, 30 and
65\% Ca, and the end compounds. The main edge is included above each
set of traces (multiplied by 0.03 to fit on the graph) to show where
the structure is located relative to the edge. Note the inversion of the
structure for the CO sample (T$_{CO}$ = 270K) (b), compared to  the CMR
samples (c) (21\% Ca, T$_c$ $\sim$ 210K) and d) (30\% Ca, T$_c$ $\sim$ 260K).
}
\label{diff_spec}
\end{figure*}

\begin{figure}
\caption{A comparison of the data for the 65\% Ca sample and a simulation
obtained from a 35-65\% weighted sum of the LaMnO$_3$ and CaMnO$_3$ end
compounds. The experimental edge is significantly sharper.
}
\label{sim_65}
\end{figure}

\begin{figure}
\caption{
A plot of $\sigma^2_{CO}(T)$ for the Mn-O bond as a function of T (65\% Ca sample).
$\sigma^2_{CO}(T)$ is extracted from the $\sigma^2_{Mn-O}$(T) data presented in
Ref. {\protect \onlinecite{Booth98b}}. T$_{CO}$ $\sim$ 270K.
}
\label{D_s2_CO}
\end{figure}

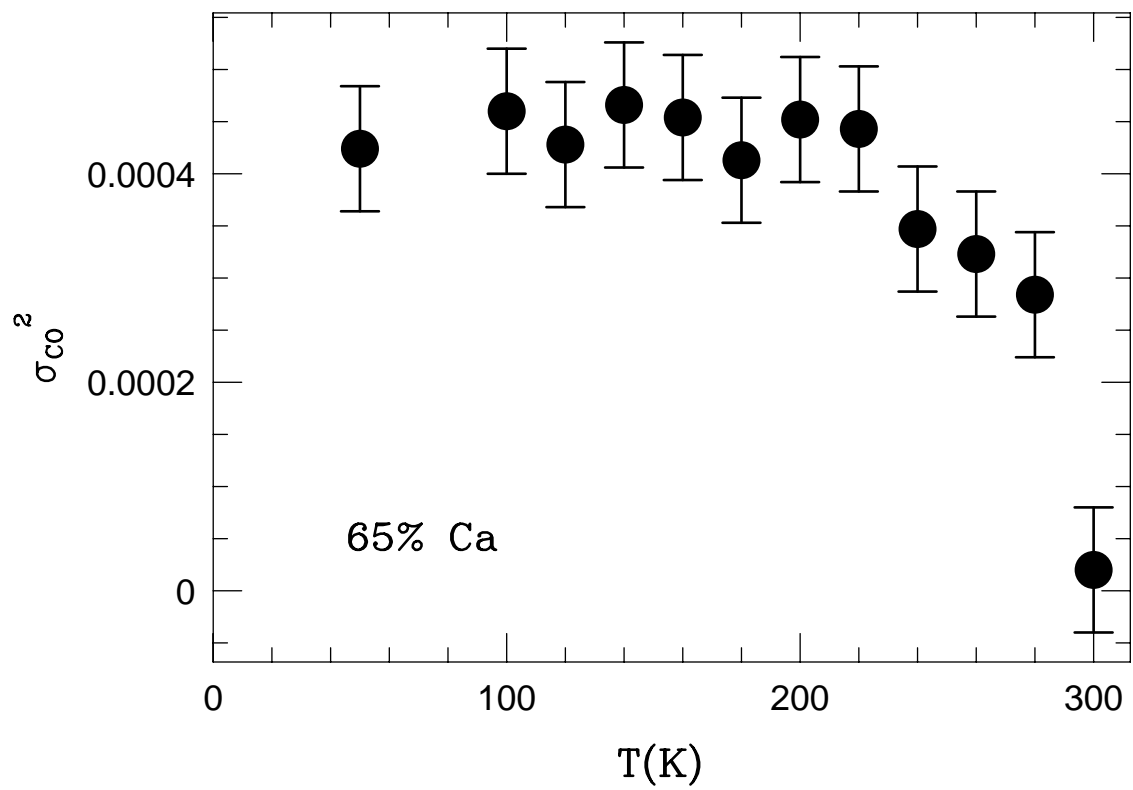
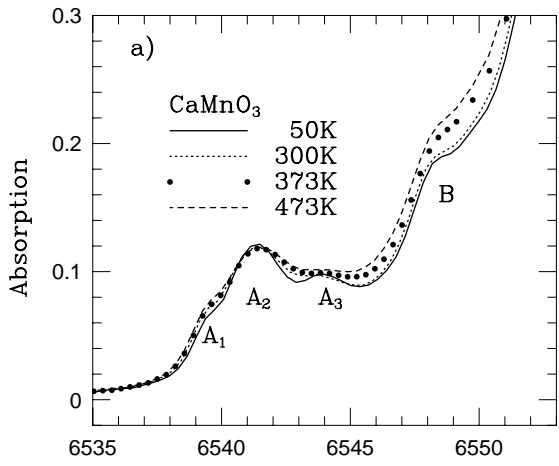
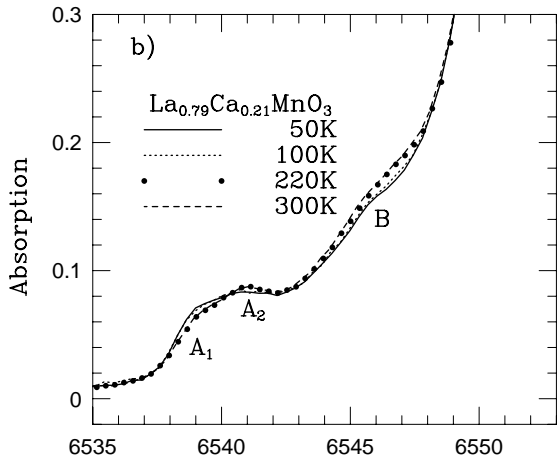
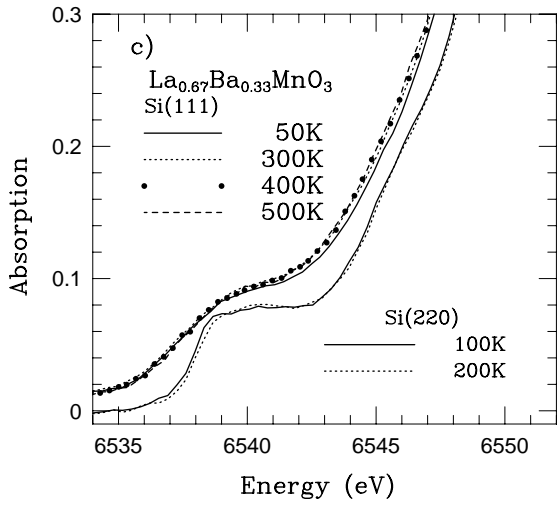
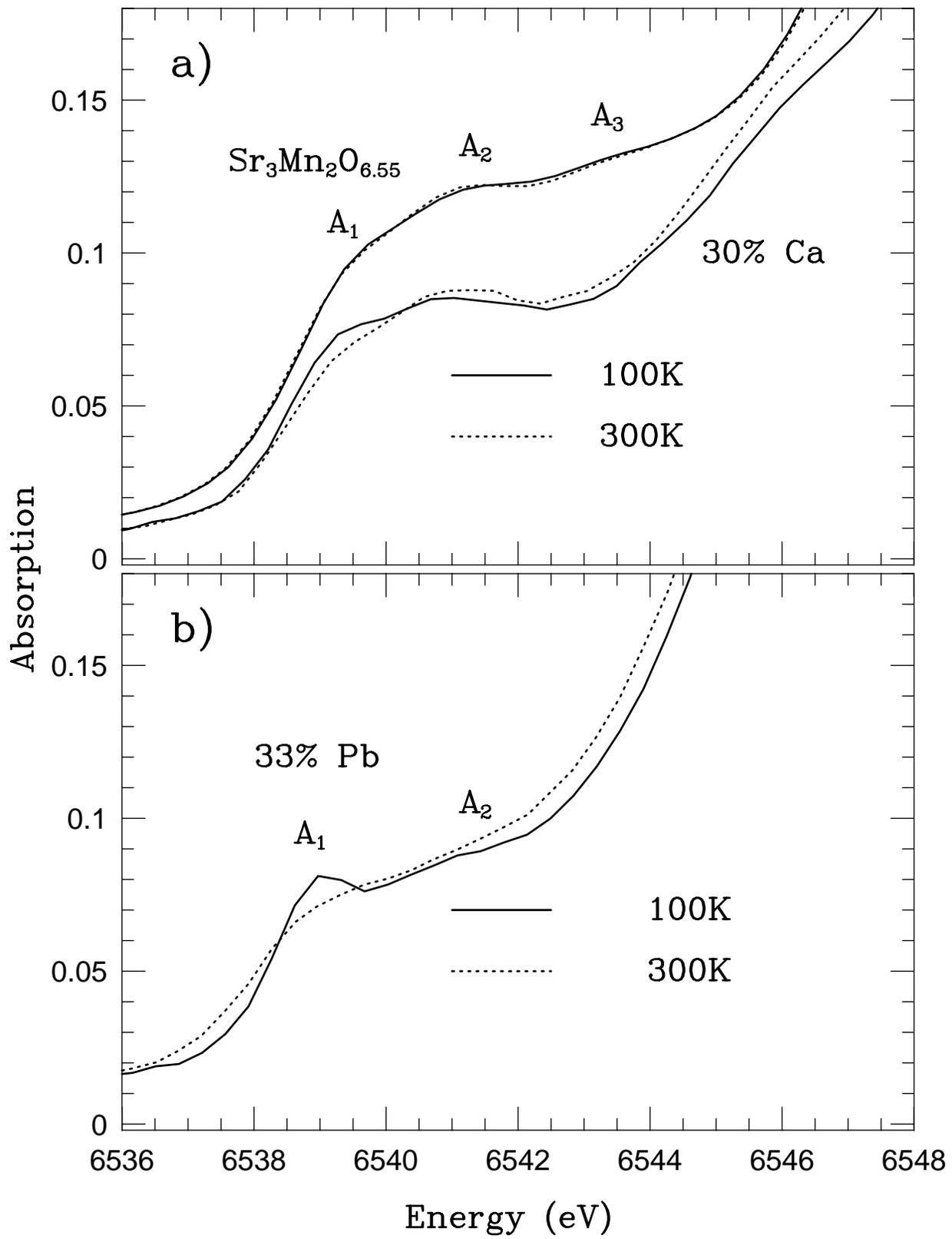
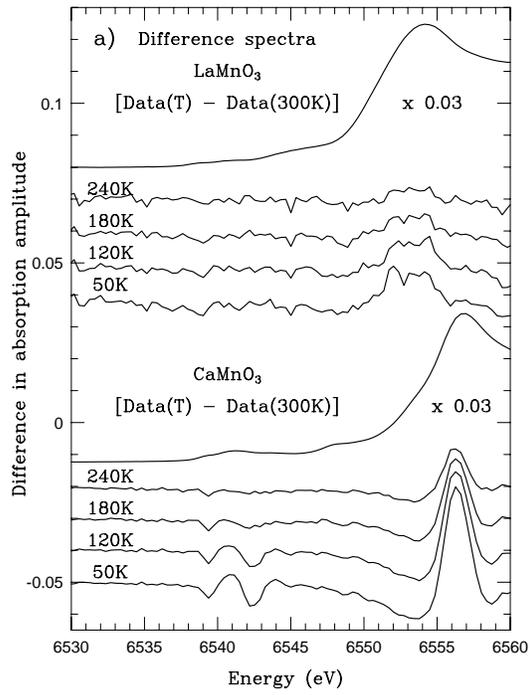
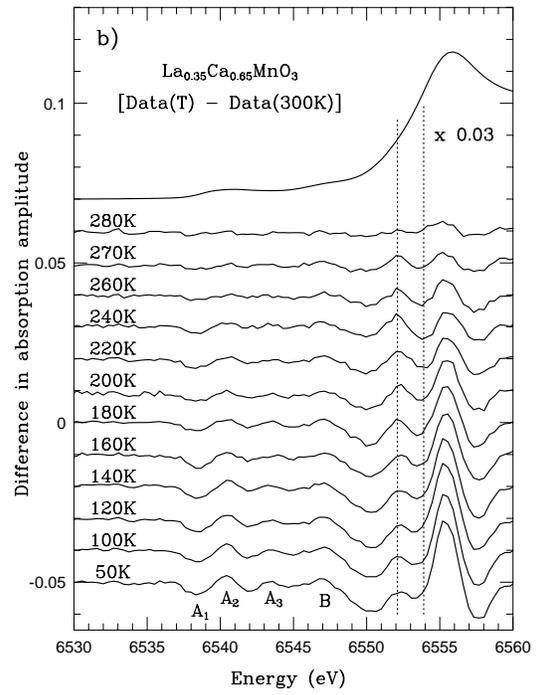
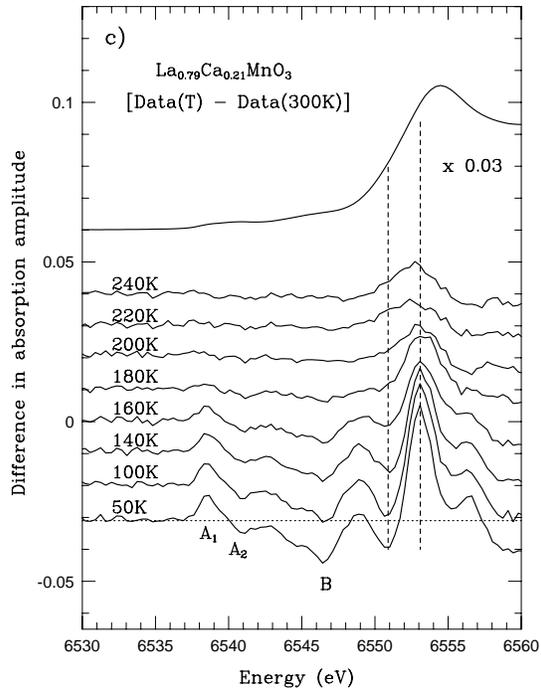
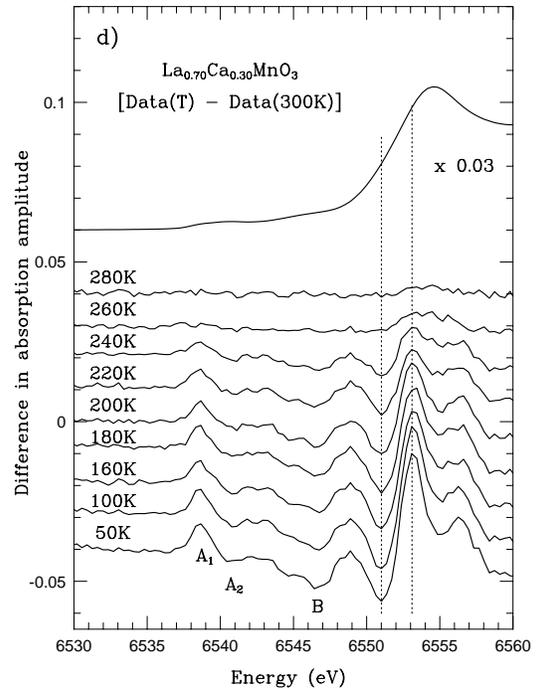
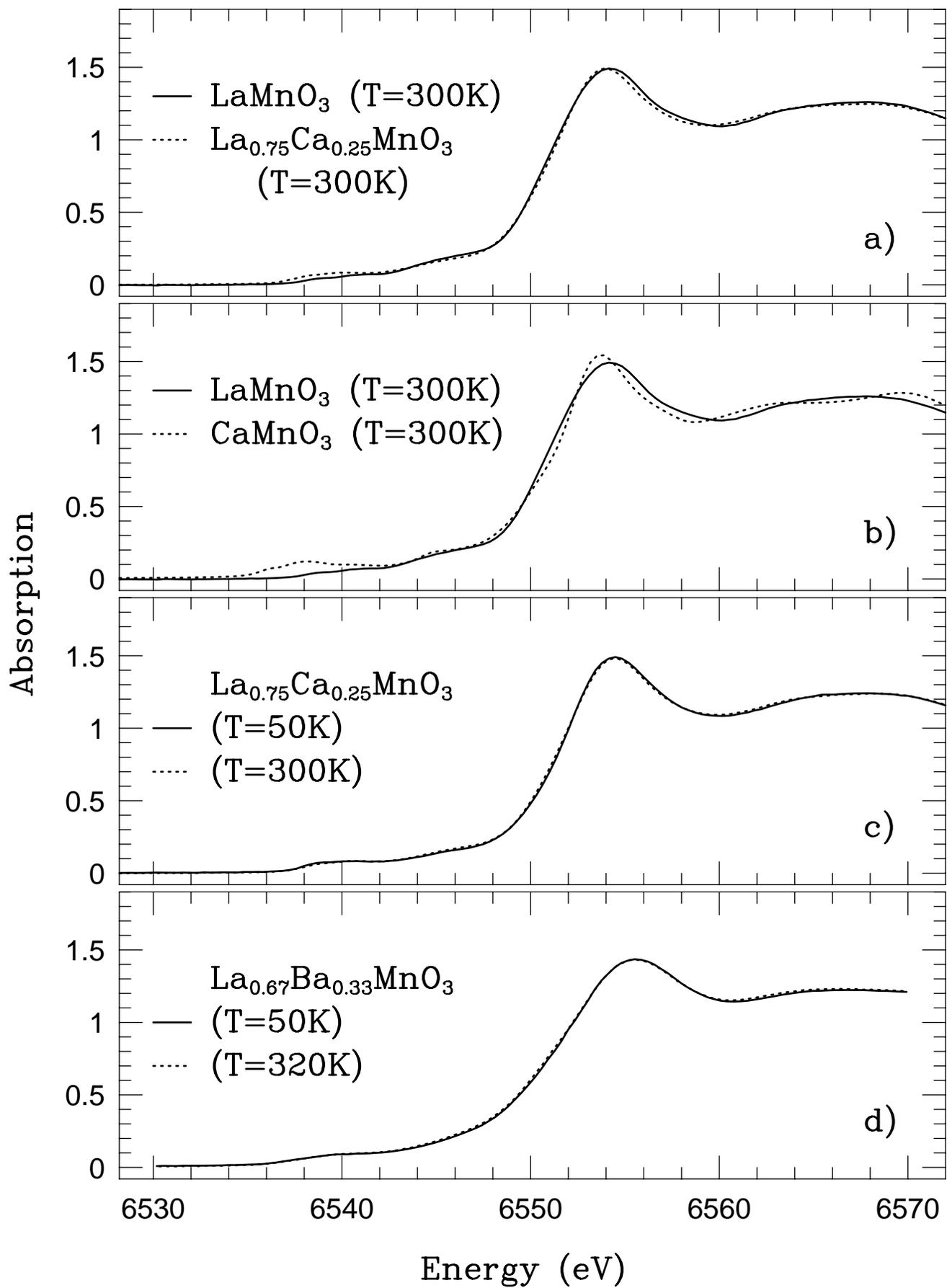
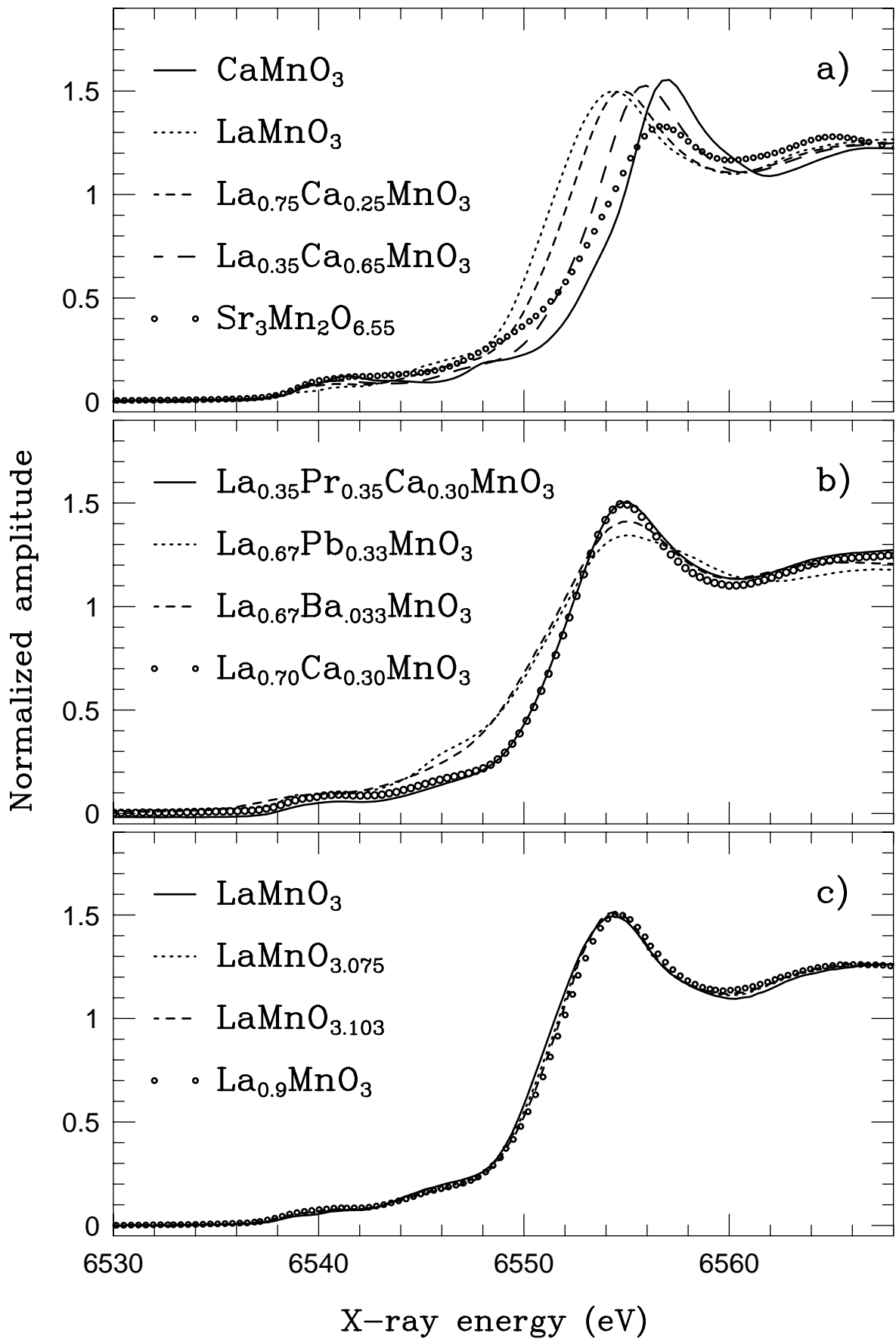
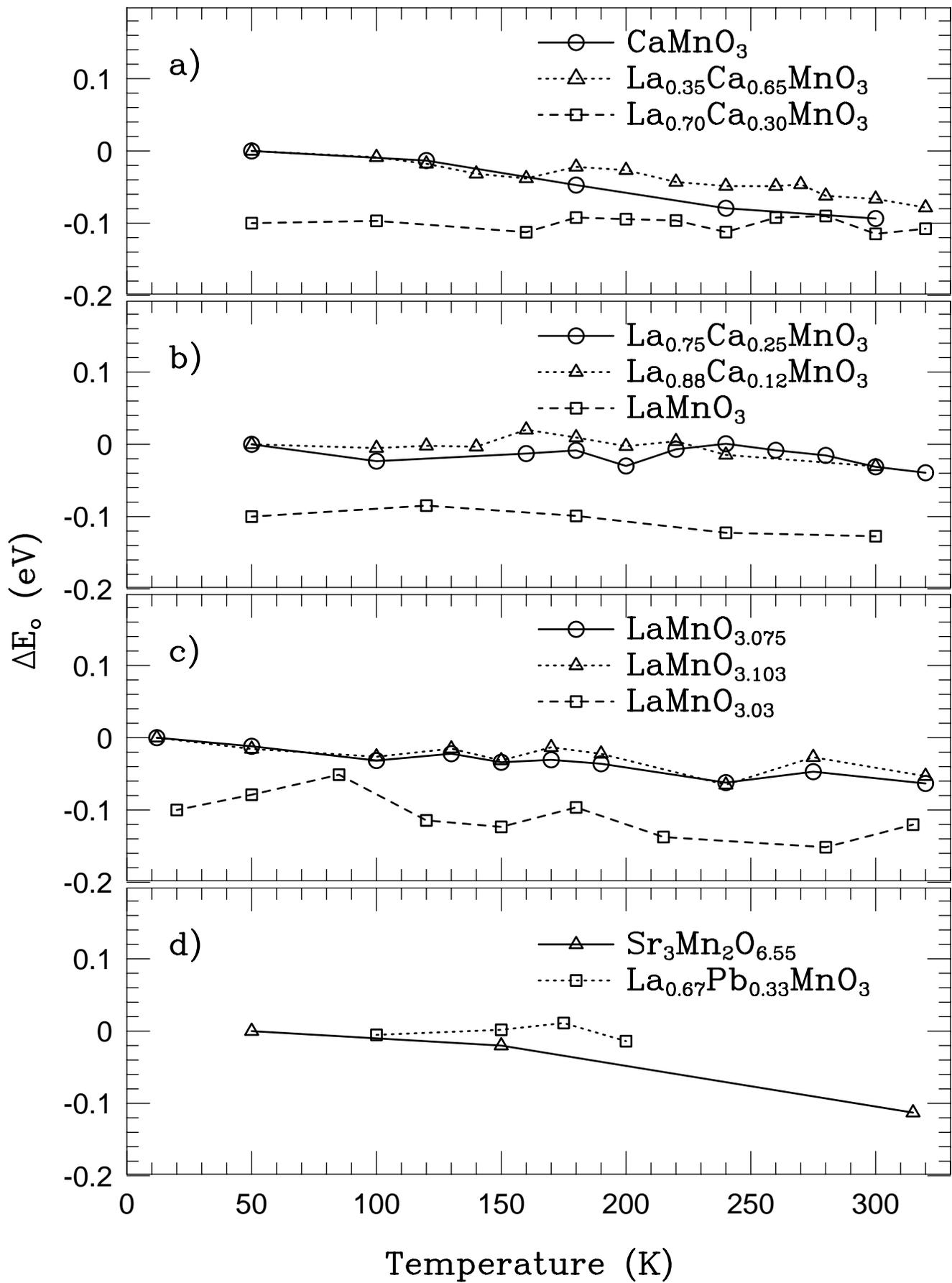
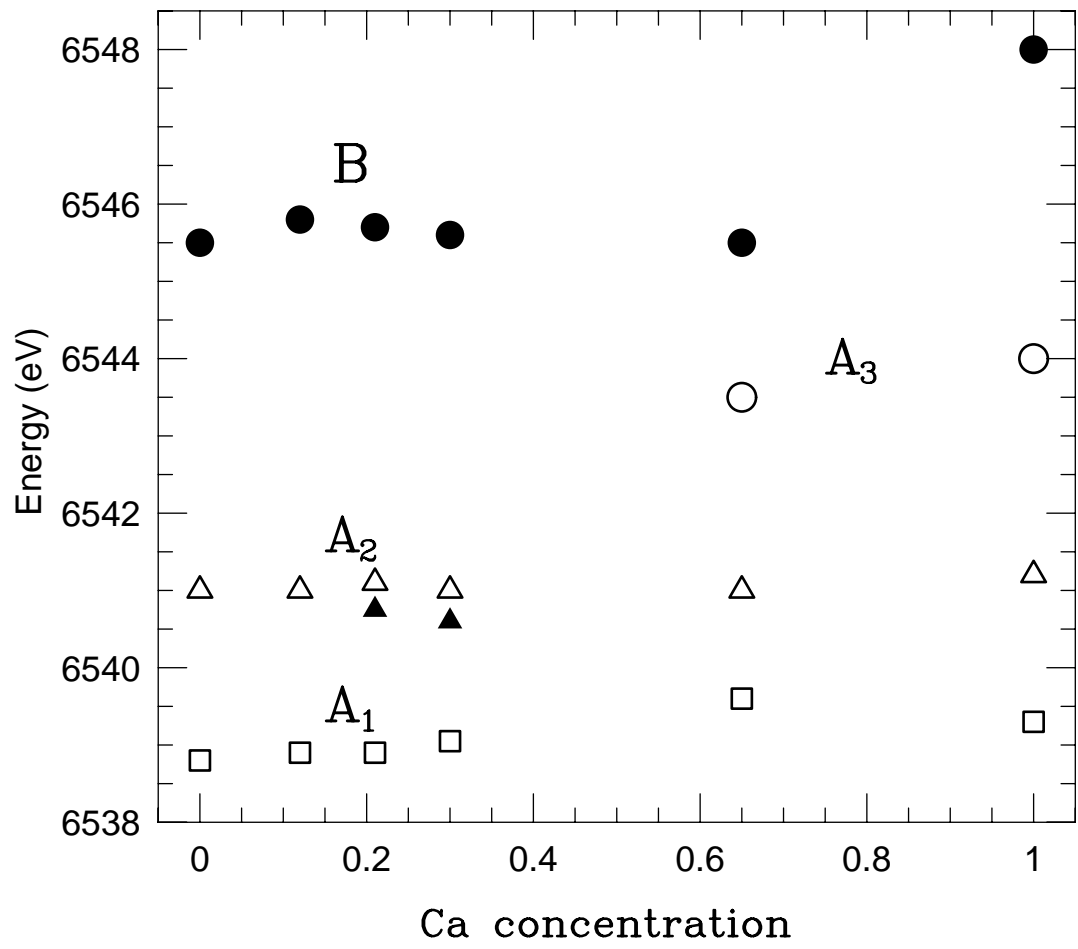
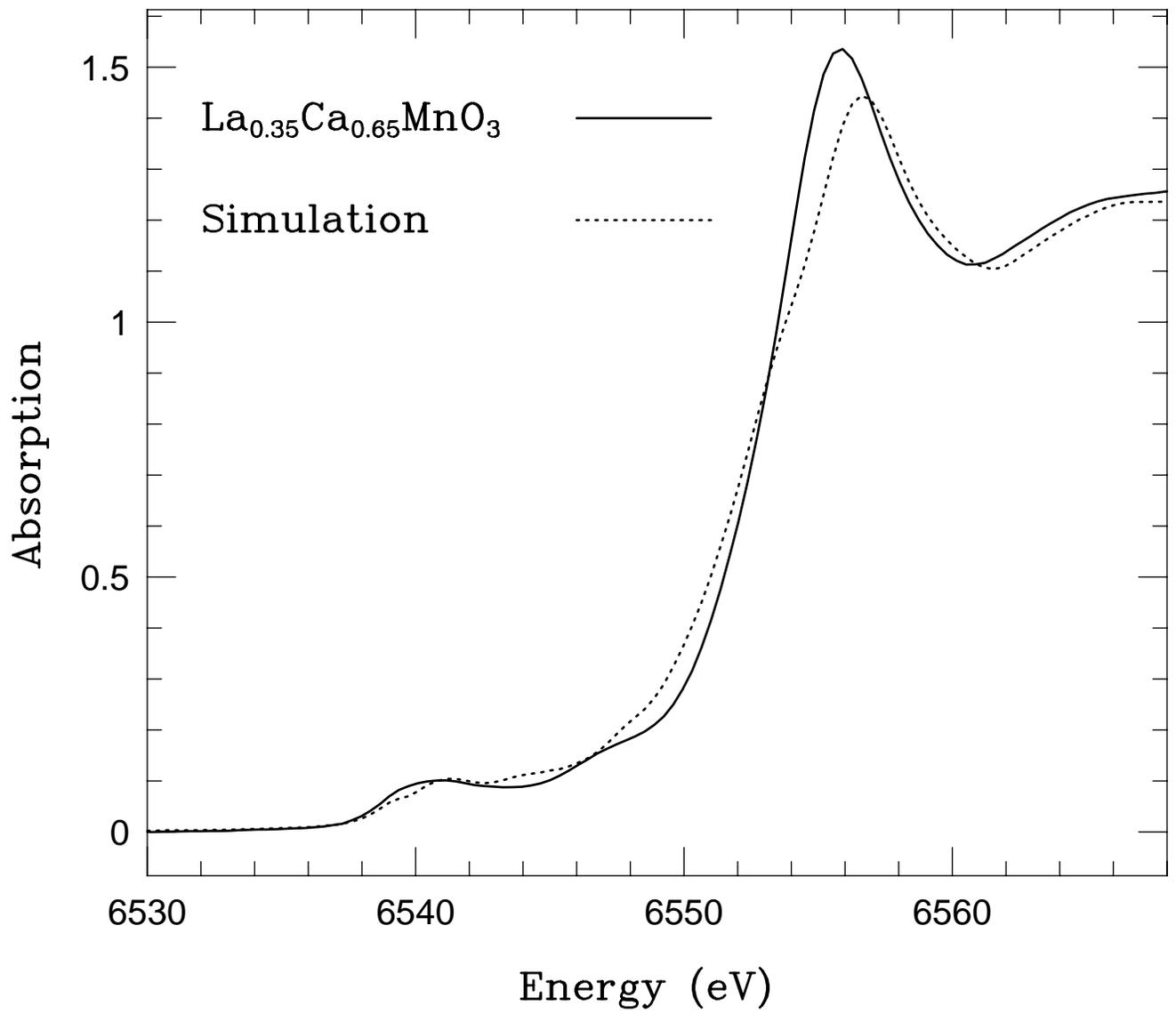
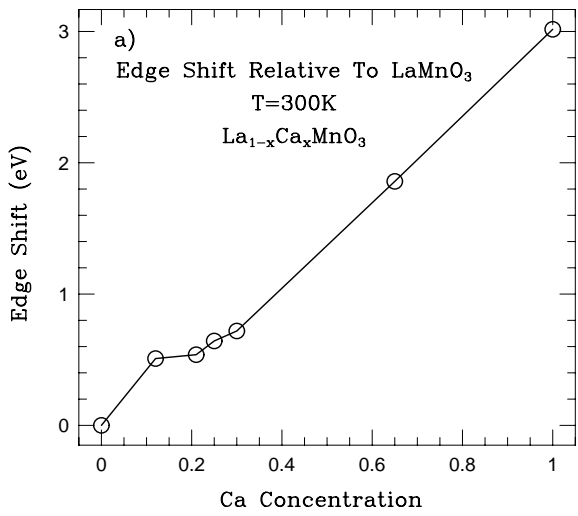
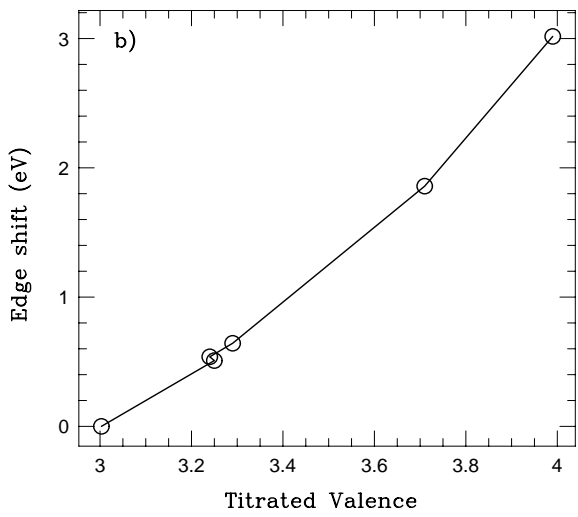
\end{document}